\documentstyle[12pt,qqa4lart]{article}
\input tcilatex

\QQQ{Language}{
American English
}

\begin{document}

\author{Lu-Ming Duan$^{*}$ and Guang-Can Guo\thanks{%
E-mail: gcguo@sunlx06.nsc.ustc.edu.cn} \\
%EndAName
Department of Physics, University of Science \\
and Technology of China, Hefei, 230026, P.R.China}
\title{Two non-orthogonal states can be cloned by a unitary-reduction process}
\date{}
\maketitle

\begin{abstract}
\baselineskip 18ptWe show that, there are physical means for cloning two
non-orthogonal pure states which are secretly chosen from a certain set $%
\$=\left\{ \left| \Psi _0\right\rangle ,\left| \Psi _1\right\rangle \right\} 
$. The states are cloned through a unitary evolution together with a
measurement. The cloning efficiency can not attain $100\%$. With some
negative measurement results, the cloning fails.\\

\end{abstract}

\newpage\ \baselineskip 18ptThe development of quantum information theory
[1] draws attention to fundamental questions about what is physically
possible and what is not. An example is the quantum no-cloning theorem [2],
which asserts, unknown pure states can not be reproduced or copied by any
physical means. Recently, there are growing interests in the no-cloning
theorem. The original proof of this theorem [2] shows that the cloning
machine violates the quantum superposition principle, which applies to a
minimum total number of three states, and hence does not rule out the
possibility of cloning two non-orthogonal states. Refs. [3] and [4] show
that a violation of unitarity makes cloning two non-orthogonal states
impossible. The result has also been extended to mixed states. That is the
quantum no-broadcasting theorem [5], which states, two non-commuting mixed
states can not be broadcast onto two separate quantum systems, even when the
states need only be reproduced marginally. With the fact that quantum states
can not be cloned ideally, recently, inaccurate copying of quantum states
arouse great interests [6-9].

In this letter, we show that, however, there are physical means for cloning
two non-orthogonal pure states. This does not contradict to the previous
proofs of the no-cloning or no-broadcasting theorem. The proof in [2]
applies to at least three states. Though the no-cloning theorem was extended
to two states in [3] and [4], the proof only holds for the unitary
evolution, not for any physical means. In particular, measurements are not
considered. Similarly, the no-broadcasting theorem proven in [5] is also
limited to the unitary evolution. ( This time it is a generalized unitary
evolution by introducing an ancillary system. ) To show this, we note in the
proof the inequality 
\begin{equation}
\label{1}F\left( \rho _0,\rho _1\right) \leq F\left( \widetilde{\rho }_0,%
\widetilde{\rho }_1\right) 
\end{equation}
plays an essential role, ( Eq. (17) in Ref. [5] ), where $\rho _s$ and $%
\widetilde{\rho }_s$ $\left( s=0,1\right) $ are the density operators before
and after the evolution, respectively. $F$ indicates the fidelity, which is
defined by 
\begin{equation}
\label{2}F\left( \rho _0,\rho _1\right) =tr\sqrt{\sqrt{\rho _0}\rho _1\sqrt{%
\rho _0}}. 
\end{equation}
In Ref. [5], the inequality (1) was proven for the general unitary
evolution. Though there is a theorem in [10], which states, the fidelity (2)
does not decrease through measurements, the proof exclude the ''read-out''
(or the projection) step. So the evolution there is still a general unitary
evolution, not a real measurement. In fact, the inequality (1) is not true
for the measurement process. We show it by the following example.

Consider two pure states $\left| \Psi _0\right\rangle $ and $\left| \Psi
_1\right\rangle $, which are defined by 
\begin{equation}
\label{3}
\begin{array}{c}
\left| \Psi _0\right\rangle =\frac 1{
\sqrt{2}}\left( \left| s_1\right\rangle +\left| s_3\right\rangle \right) ,
\\  \\ 
\left| \Psi _1\right\rangle =\frac 1{\sqrt{2}}\left( \left| s_2\right\rangle
+\left| s_3\right\rangle \right) ,
\end{array}
\end{equation}
where $\left| s_1\right\rangle $, $\left| s_2\right\rangle $, and $\left|
s_3\right\rangle $ are three eigenstates of an observable $S$ with the
eigenvalues $s_1$, $s_2$, and $s_3$, respectively. We measure the observable 
$S$. If the measurement result is $s_3$, the output state is discarded. With
the input states $\left| \Psi _0\right\rangle $ and $\left| \Psi
_1\right\rangle $, the output states are respectively 
\begin{equation}
\label{4}
\begin{array}{c}
\left| 
\widetilde{\Psi }_0\right\rangle =\left| s_1\right\rangle , \\  \\ 
\left| \widetilde{\Psi }_1\right\rangle =\left| s_2\right\rangle .
\end{array}
\end{equation}
It is obvious that 
\begin{equation}
\label{5}F\left( \rho _0,\rho _1\right) =\left| \left\langle \Psi _0|\Psi
_1\right\rangle \right| =\frac 12>0=F\left( \widetilde{\rho }_0,\widetilde{%
\rho }_1\right) .
\end{equation}
So the inequality (1) does not hold for the measurement process.

Now we prove that two non-orthogonal state can be cloned by a unitary
evolution together with a measurement. The result, posed formally, is the
following theorem

{\it Theorem.} If $\left| \Psi _0\right\rangle $ and $\left| \Psi
_1\right\rangle $ are two non-orthogonal states of a quantum system A ,there
exist a unitary evolution $U$ and a measurement $M$, which together yield
the following evolution 
\begin{equation}
\label{6}
\begin{array}{c}
\left| \Psi _0\right\rangle \left| \Sigma \right\rangle 
\stackrel{U+M}{\longrightarrow }\left| \Psi _0\right\rangle \left| \Psi
_0\right\rangle , \\  \\ 
\left| \Psi _1\right\rangle \left| \Sigma \right\rangle \stackrel{U+M}{%
\longrightarrow }\left| \Psi _1\right\rangle \left| \Psi _1\right\rangle ,
\end{array}
\end{equation}
where $\left| \Sigma \right\rangle $ is the input state of a system B.
Systems A and B each have an $n$-dimensional Hilbert space.

{\it Proof.} First we consider the measurement. If there exists a unitary
operator $U$ to make 
\begin{equation}
\label{7}
\begin{array}{c}
U\left( \left| \Psi _0\right\rangle \left| \Sigma \right\rangle \left|
m_0\right\rangle \right) =a_{00}\left| \Psi _0\right\rangle \left| \Psi
_0\right\rangle \left| m_0\right\rangle +a_{01}\left| \Phi
_{AB}\right\rangle \left| m_1\right\rangle , \\  
\\ 
U\left( \left| \Psi _1\right\rangle \left| \Sigma \right\rangle \left|
m_0\right\rangle \right) =a_{10}\left| \Psi _1\right\rangle \left| \Psi
_1\right\rangle \left| m_0\right\rangle +a_{11}\left| \Phi
_{AB}\right\rangle \left| m_1\right\rangle ,
\end{array}
\end{equation}
where $\left| m_0\right\rangle $ and $\left| m_1\right\rangle $ are two
orthogonal states of a probe P, in succession we measure the probe P, and
the states are preserved if the measurement result is $m_0$. This
measurement projects the composite system AB into the state $\left| \Psi
_s\right\rangle \left| \Psi _s\right\rangle $, where $s=0$ or $1$. So the
evolution (6) exists if Eq. (7) holds. To prove existence of the unitary
operator $U$ described by Eq. (7), we first introduce two lemmas.

{\it Lemma 1.} If the normalized states $\left| \phi _0\right\rangle $, $%
\left| \phi _1\right\rangle $, $\left| \widetilde{\phi }_0\right\rangle $,
and $\left| \widetilde{\phi }_1\right\rangle $ satisfy $\left\langle \phi
_0|\phi _1\right\rangle =\left\langle \widetilde{\phi }_0|\widetilde{\phi }%
_1\right\rangle =0$, there exists a unitary operator $U$ to make 
\begin{equation}
\label{8}
\begin{array}{c}
U\left| \phi _0\right\rangle =\left| 
\widetilde{\phi }_0\right\rangle , \\  \\ 
U\left| \phi _1\right\rangle =\left| \widetilde{\phi }_1\right\rangle . 
\end{array}
\end{equation}

{\it Proof:} Suppose the considered system has an $n$-dimensional Hilbert
space $H$. There exist $n-2$ orthogonal states $\left| \phi _2\right\rangle $%
, $\left| \phi _3\right\rangle $, $\cdots $, $\left| \phi
_{n-1}\right\rangle ,$ which together with $\left| \phi _0\right\rangle $
and $\left| \phi _1\right\rangle $ make an orthonormal basis for the space $%
H $. Similarly, the states $\left| \widetilde{\phi }_0\right\rangle $, $%
\left| \widetilde{\phi }_1\right\rangle $, $\cdots ,$ and $\left| \widetilde{%
\phi }_{n-1}\right\rangle $ make another orthonormal basis. The following
operator 
\begin{equation}
\label{9}U=\left| \widetilde{\phi }_0\right\rangle \left\langle \phi
_0\right| +\left| \widetilde{\phi }_1\right\rangle \left\langle \phi
_1\right| +\cdots +\left| \widetilde{\phi }_{n-1}\right\rangle \left\langle
\phi _{n-1}\right| 
\end{equation}
is unitary, which can be easily checked by verifying the identity 
\begin{equation}
\label{10}UU^{+}=U^{+}U=I. 
\end{equation}
The operator $U$ defined by Eq. (9) evolves the states $\left| \phi
_0\right\rangle $ and $\left| \phi _1\right\rangle $ into $\left| \widetilde{%
\phi }_0\right\rangle $ and $\left| \widetilde{\phi }_1\right\rangle $,
respectively. Lemma 1 is thus proved.

{\it Lemma 2.} If the states $\left| \phi _0\right\rangle $, $\left| \phi
_1\right\rangle $, $\left| \widetilde{\phi }_0\right\rangle ,$ and $\left| 
\widetilde{\phi }_1\right\rangle $ satisfy 
\begin{equation}
\label{11}
\begin{array}{c}
\left\langle \phi _0|\phi _0\right\rangle =\left\langle 
\widetilde{\phi }_0|\widetilde{\phi }_0\right\rangle , \\  \\ 
\left\langle \phi _1|\phi _1\right\rangle =\left\langle 
\widetilde{\phi }_1|\widetilde{\phi }_1\right\rangle , \\  \\ 
\left\langle \phi _0|\phi _1\right\rangle =\left\langle \widetilde{\phi }_0|%
\widetilde{\phi }_1\right\rangle , 
\end{array}
\end{equation}
there exists a unitary operator $U$ to make $U\left| \phi _0\right\rangle
=\left| \widetilde{\phi }_0\right\rangle ,$ and $U\left| \phi
_1\right\rangle =\left| \widetilde{\phi }_1\right\rangle $.

{\it Proof:} Suppose $\gamma _0=\left\| \left| \phi _0\right\rangle \right\| 
$, and $\gamma _1=\left\| \left| \phi _1\right\rangle -\frac{\left\langle
\phi _0|\phi _1\right\rangle }{\gamma _0^2}\left| \phi _0\right\rangle
\right\| $, where the norm $\left\| \left| \phi \right\rangle \right\| $ is
defined by $\left\| \left| \phi \right\rangle \right\| =\sqrt{\left\langle
\phi |\phi \right\rangle }$. The normalized states $\frac 1{\gamma _0}\left|
\phi _0\right\rangle $ and\\ $\frac 1{\gamma _1}\left( \left| \phi
_1\right\rangle -\frac{\left\langle \phi _0|\phi _1\right\rangle }{\gamma
_0^2}\left| \phi _0\right\rangle \right) $ are obviously orthogonal. On the
other hand, following Eq. (11), the two states $\frac 1{\gamma _0}\left| 
\widetilde{\phi }_0\right\rangle $ and $\frac 1{\gamma _1}\left( \left| 
\widetilde{\phi }_1\right\rangle -\frac{\left\langle \phi _0|\phi
_1\right\rangle }{\gamma _0^2}\left| \widetilde{\phi }_0\right\rangle
\right) $ are also normalized and orthogonal. Hence, from Lemma 1 , there
exists a unitary operator $U$ to make 
\begin{equation}
\label{12}
\begin{array}{c}
U\left( \frac 1{\gamma _0}\left| \phi _0\right\rangle \right) =\frac
1{\gamma _0}\left| 
\widetilde{\phi }_0\right\rangle , \\  \\ 
U\left( \frac 1{\gamma _1}\left( \left| \phi _1\right\rangle -\frac{%
\left\langle \phi _0|\phi _1\right\rangle }{\gamma _0^2}\left| \phi
_0\right\rangle \right) \right) =\frac 1{\gamma _1}\left( \left| \widetilde{%
\phi }_1\right\rangle -\frac{\left\langle \phi _0|\phi _1\right\rangle }{%
\gamma _0^2}\left| \widetilde{\phi }_0\right\rangle \right) .
\end{array}
\end{equation}
Eq. (12) is just another expression of the evolution $U\left| \phi
_0\right\rangle =\left| \widetilde{\phi }_0\right\rangle $ and $U\left| \phi
_1\right\rangle =\left| \widetilde{\phi }_1\right\rangle $. Lemma 2 is
therefore proved.

Now we return to the proof of the main theorem. Let 
\begin{equation}
\label{13}
\begin{array}{c}
\left| \phi _s\right\rangle =\left| \Psi _s\right\rangle \left| \Sigma
\right\rangle \left| m_0\right\rangle , \\  
\\ 
\left| \widetilde{\phi }_s\right\rangle =a_{s0}\left| \Psi _s\right\rangle
\left| \Psi _s\right\rangle \left| m_0\right\rangle +a_{s1}\left| \Phi
_{AB}\right\rangle \left| m_1\right\rangle , 
\end{array}
\end{equation}
where $s=0$ or $1$. Without loss of generality, here and in the following we
suppose $\left\langle \Psi _0|\Psi _1\right\rangle $ is a positive real
number. It can be easily checked that, if the constants $a_{00}$, $a_{01}$, $%
a_{10}$, and $a_{11}$ in Eq. (7) have the following values 
\begin{equation}
\label{14}
\begin{array}{c}
a_{00}=a_{10}=\frac 1{
\sqrt{1+\left\langle \Psi _0|\Psi _1\right\rangle }}, \\  \\ 
a_{01}=a_{11}=\frac{\sqrt{\left\langle \Psi _0|\Psi _1\right\rangle }}{\sqrt{%
1+\left\langle \Psi _0|\Psi _1\right\rangle }}, 
\end{array}
\end{equation}
the states $\left| \phi _0\right\rangle $, $\left| \phi _1\right\rangle $, $%
\left| \widetilde{\phi }_0\right\rangle $, and $\left| \widetilde{\phi }%
_1\right\rangle $ defined by Eq. (13) satisfy the condition (11). So there
exists a unitary operator $U$ to realize the evolution (7). This completes
the proof of the main theorem.

The above proof of the theorem is constructive, i.e., it gives a method for
constructing the desired unitary operator $U$ and the measurement $M$. We
illustrate the construction by the following example.

{\it Example.} We consider the simplest system which consists of three
parts, A and B and a probe P, each being a qubit ( a two-state quantum
system ). We want to clone two non-orthogonal states $\left| \Psi
_0\right\rangle $ and $\left| \Psi _1\right\rangle $ of the qubit A with $%
\left\langle \Psi _0|\Psi _1\right\rangle =\cos \left( \theta \right) =\tan
^2\alpha $, where $0\leq \alpha <\frac \pi 4$. For this system, the states
in Eq. (7) can be chosen as $\left| \Psi _0\right\rangle =\left|
0\right\rangle $, $\left| \Psi _1\right\rangle =\cos \left( \theta \right)
\left| 0\right\rangle +\sin \left( \theta \right) \left| 1\right\rangle $, $%
\left| \Sigma \right\rangle =\left| 0\right\rangle $, $\left|
m_0\right\rangle =\left| 0\right\rangle $, $\left| m_1\right\rangle =\left|
1\right\rangle $, $\left| \Phi _{AB}\right\rangle =\left| 00\right\rangle .$
We measure the qubit P. With the measurement result $0$, the states of the
qubit A are successfully cloned. According to Eq. (9), the unitary operator $%
U$ has the form $U=\stackrel{7}{\stackunder{i=0}{\Sigma }}\left| \widetilde{%
\phi }_i\right\rangle \left\langle \phi _i\right| $. A natural choice of the
states $\left| \phi _i\right\rangle $ is the computational basis $\left|
000\right\rangle $, $\left| 100\right\rangle $, $\cdots $, $\left|
111\right\rangle $. From Eq. (14) and the proof of Lemma 2, the states $%
\left| \widetilde{\phi }_0\right\rangle $ and $\left| \widetilde{\phi }%
_1\right\rangle $ are respectively 
\begin{equation}
\label{15}
\begin{array}{c}
\left| 
\widetilde{\phi }_0\right\rangle =\cos \alpha \left| 000\right\rangle +\sin
\alpha \left| 001\right\rangle , \\  \\ 
\left| 
\widetilde{\phi }_1\right\rangle =-\sqrt{\cos \left( 2\alpha \right) }\sin
\alpha \tan \alpha \left| 000\right\rangle +\sin \alpha \tan \alpha \left(
\left| 100\right\rangle +\left| 010\right\rangle \right) \\  \\ 
+\sqrt{1-\tan ^2\alpha }\left| 110\right\rangle +\sqrt{\cos \left( 2\alpha
\right) }\sin \alpha \left| 001\right\rangle . 
\end{array}
\end{equation}
$\left| \widetilde{\phi }_0\right\rangle $ and $\left| \widetilde{\phi }%
_1\right\rangle $ are superpositions of the states $\left| \phi
_0\right\rangle $, $\left| \phi _1\right\rangle $, $\cdots $, and $\left|
\phi _4\right\rangle $, so the three states $\left| \widetilde{\phi }%
_5\right\rangle $, $\left| \widetilde{\phi }_6\right\rangle ,$ and $\left| 
\widetilde{\phi }_7\right\rangle $ can be chosen as $\left| \phi
_5\right\rangle $, $\left| \phi _6\right\rangle ,$ $\left| \phi
_7\right\rangle $, respectively. The states $\left| \widetilde{\phi }%
_2\right\rangle $, $\left| \widetilde{\phi }_3\right\rangle ,$ and $\left| 
\widetilde{\phi }_4\right\rangle $ are also superpositions of the states $%
\left| \phi _0\right\rangle $, $\left| \phi _1\right\rangle $, $\cdots $, $%
\left| \phi _4\right\rangle $. The superposition constants are determined by
the orthonormal conditions. This evolution leaves the subspace spanned by $%
\left| \phi _5\right\rangle $, $\left| \phi _6\right\rangle ,$ and $\left|
\phi _7\right\rangle $ unchanged, and makes a rotation in the subspace
spanned by $\left| \phi _0\right\rangle $, $\left| \phi _1\right\rangle $, $%
\cdots $, $\left| \phi _4\right\rangle $. The evolution $U$ can be fulfilled
by the quantum controlled-NOT gates together with some single-qubit rotation
gates [11,12]. From this example, we see, even for the simplest system, the
evolution yielding the cloning is rather complicated.

At the first glance, that two non-orthogonal states can be cloned seemingly
threatens the security of the quantum cryptography schemes based on
non-orthogonal states [13-15]. But this is not the case. The key reason is
that, though two non-orthogonal states can be cloned, they can not be cloned
always successfully. If the measurement of the probe does not yield the
desired result $m_0$, the cloning fails. Through these failures, Alice (the
sender) and Bob ( the receiver) can find the intervention of Eve (the
eavesdropper).

In Eq. (7), with probability $\eta _0=a_{00}^2$ and $\eta _1=a_{10}^2$, the
measurement of the probe yields the desired cloned states for the composite
system AB. So $\eta _0$ and $\eta _1$ define the cloning efficiencies. Now
we prove that, for any cloning machines, the cloning efficiencies can not
attain $100\%$ at the same time. They must satisfy some basic inequalities.

A general unitary transformation of pure states can be decomposed as 
\begin{equation}
\label{16}
\begin{array}{c}
U\left( \left| \Psi _0\right\rangle \left| \Sigma \right\rangle \left|
m_p\right\rangle \right) =
\sqrt{\eta _0}\left| \Psi _0\right\rangle \left| \Psi _0\right\rangle \left|
m_0\right\rangle +\sqrt{1-\eta _0}\left| \Phi _{ABP}^0\right\rangle , \\  \\ 
U\left( \left| \Psi _1\right\rangle \left| \Sigma \right\rangle \left|
m_p\right\rangle \right) =\sqrt{\eta _1}\left| \Psi _1\right\rangle \left|
\Psi _1\right\rangle \left| m_1\right\rangle +\sqrt{1-\eta _1}\left| \Phi
_{ABP}^1\right\rangle ,
\end{array}
\end{equation}
where $\left| m_p\right\rangle $, $\left| m_0\right\rangle ,$ and $\left|
m_1\right\rangle $ are states of the probe, and $\left| \Phi
_{ABP}^0\right\rangle $ and $\left| \Phi _{ABP}^1\right\rangle $ are two
states of the composite system ABP. In general, they are not necessarily
orthogonal to each other. In the cloning, a measurement projects the states
of the probe into the subspace spanned by $\left| m_0\right\rangle $ and $%
\left| m_1\right\rangle $. After projection, the state of the system AB
should be $\left| \Psi _s\right\rangle \left| \Psi _s\right\rangle $, where $%
s=0$ or $1$. This requires 
\begin{equation}
\label{17}\left\langle m_0|\Phi _{ABP}^0\right\rangle =\left\langle m_1|\Phi
_{ABP}^0\right\rangle =\left\langle m_0|\Phi _{ABP}^1\right\rangle
=\left\langle m_1|\Phi _{ABP}^1\right\rangle =0.
\end{equation}
The above condition suggests that the cloning here is different from the
inaccurate quantum copying in Ref. [7], where one does not need to require $%
\left\langle m_1|\Phi _{ABP}^0\right\rangle =\left\langle m_0|\Phi
_{ABP}^1\right\rangle =0.$ With the condition (17), inner product of the two
parts of Eq. (16) gives 
\begin{equation}
\label{18}
\begin{array}{c}
\left\langle \Psi _0|\Psi _1\right\rangle -
\sqrt{\eta _0\eta _1}\left\langle \Psi _0|\Psi _1\right\rangle
^2\left\langle m_0|m_1\right\rangle  \\  \\ 
=
\sqrt{\left( 1-\eta _0\right) \left( 1-\eta _1\right) }\left\langle \Phi
_{ABP}^0|\Phi _{ABP}^1\right\rangle  \\  \\ 
\leq \sqrt{\left( 1-\eta _0\right) \left( 1-\eta _1\right) }.
\end{array}
\end{equation}
In Eq. (18), we use the inequality $\left| \left\langle \Phi _{ABP}^0|\Phi
_{ABP}^1\right\rangle \right| \leq \left\| \left| \Phi _{ABP}^0\right\rangle
\right\| \left\| \left| \Phi _{ABP}^1\right\rangle \right\| $. From Eq. (18)
, it is not difficult to obtain that 
\begin{equation}
\label{19}\frac{\eta _0+\eta _1}2\leq \frac{1-\left\langle \Psi _0|\Psi
_1\right\rangle }{1-\left\langle \Psi _0|\Psi _1\right\rangle ^2\left\langle
m_0|m_1\right\rangle }\leq \frac 1{1+\left\langle \Psi _0|\Psi
_1\right\rangle }.
\end{equation}
We have supposed $\left\langle \Psi _0|\Psi _1\right\rangle \neq 1$. This
inequality suggests that the efficiencies $\eta _0$ and $\eta _1$ can not
attain $100\%$ at the same time for two non-orthogonal states. The equality
in Eq. (19) holds if and only if $\eta _0=\eta _1$ and $\left|
m_0\right\rangle =\left| m_1\right\rangle $. The case $\eta _0=\eta _1$ is
of special interest. In this case, the cloning efficiency is independent of
the input states. Such a cloning machine is called the universal quantum
cloning machine. With $\eta _0=\eta _1=\eta $, Eq. (19) reduces to 
\begin{equation}
\label{20}\eta \leq \frac 1{1+\left\langle \Psi _0|\Psi _1\right\rangle }.
\end{equation}
Hence the efficiency given by Eq. (14) is in fact the maximum efficiency
able to be obtained by a cloning machine. Also, for the universal cloning
machine, the cloning efficiency is always less than $100\%$. This explains
the security of the quantum cryptography schemes based on two non-orthogonal
states [13-15]. The reason is not that non-orthogonal states can not be
cloned, but that the cloning efficiency can not attain $100\%$.\\

{\bf Acknoledgment}

This project was supported by the National Natural Science Foundation of
China.

\newpage\


\begin{thebibliography}{99}
\bibitem{1}  C.H.Bennett, Phys. Today 48, No. 10, 24 (1995).

\bibitem{2}  W.K. Wootters and W.H. Zurek, Nature (London) 299, 802 (1982).

\bibitem{3}  H.P. Yuen, Phys. Lett. A 113, 405 (1986).

\bibitem{4}  G.M. D'Ariano and H.P. Yuen, Phys. Rev. Lett. 76, 2832 (1996).

\bibitem{5}  H. Barnum, G.M. Caves, C.A. Fuchs, R. Jozsa, and B. Schumacher,
Phys. Rev. Lett. 76, 2818 (1996).

\bibitem{6}  V. Buzek and M. Hillery, Phys. Rev. A 54, 1844 (1996).

\bibitem{7}  M. Hillery and V. Buzek, Los Alamos eprint archives
quant-ph/9701034.

\bibitem{8}  V. Buzek, V.Vedral, M. Plenio, P.L. Knight, and M. Hillery, Los
Alamos eprint archives quant-ph/9701028.

\bibitem{9}  V. Buzek, S.L. Braunstein, M. Hillery, and D. Bruss, Los Alamos
eprint archives quant-ph/9703046.

\bibitem{10}  R. Jozsa, J. Mod. Opt. 41, 2315 (1994).

\bibitem{11}  A. Barenco, C.H. Bennett, R. Cleve, D.P. DiVincenzo, N.
Margolus, P. Shor, T. Sleator, J.A. Smolin, and H. Weinfurter, Phys. Rev. A
52, 3457 (1995).

\bibitem{12}  A. Barenco, D. Deutsch, A. Ekert, and R. Jozsa, Phys. Rev.
Lett. 74, 4083 (1995).

\bibitem{13}  C. H. Bennett, G. Brassard, and N.D. Mermin, Phys. Rev. Lett.
68, 557 (1992).

\bibitem{14}  C.H. Bennett, Phys. Rev. Lett. 68, 3121 (1992).

\bibitem{15}  M. Koashi and N. Imoto, Phys. Rev. Lett. 77, 2137 (1996).
\end{thebibliography}
\end{document}